\documentclass[1p]{elsarticle}
\usepackage{lineno,hyperref}
\usepackage{bm}
\usepackage{amsmath}
\usepackage{enumerate}
\usepackage{color}
\biboptions{authoryear}

\begin{document}

\begin{frontmatter}
  \title{A fast adhesive discrete element method for random packings of fine particles}


  \author{Sheng Chen}
  \author{Wenwei Liu}
  \author{Shuiqing Li\corref{mycorrespondingauthor}}
  \cortext[mycorrespondingauthor]{Corresponding author}
  \ead{lishuiqing@tsinghua.edu.cn}

  \address {Key Laboratory for Thermal Science and Power Engineering of Ministry of Education, Department of Energy and Power Engineering, Tsinghua University, Beijing, 100084, China}

  \begin{abstract}
  Introducing a reduced particle stiffness in discrete element method (DEM) allows for bigger time steps and therefore fewer total iterations in a simulation. Although this approach works well for dry non-adhesive particles, it has been shown that for fine particles with adhesion, system behaviors are drastically sensitive to the particle stiffness. Besides, a simple and applicable principle to set the parameters in adhesive DEM is also lacking. To solve these two problems, we first propose a fast DEM based on scaling laws to reduce particle Young's modulus, surface energy and to modify rolling and sliding resistances simultaneously in the framework of Johnson-Kendall-Roberts (JKR)-based contact theory. A novel inversion method is then presented to help users to quickly determine the damping coefficient, particle stiffness and surface energy to reproduce a prescribed experimental result. After validating this inversion method, we apply the fast adhesive DEM to packing problems of microparticles. Measures of packing fraction, averaged coordination number and distributions of local packing fraction and contact number of each particle are in good agreement with results simulated using original value of particle properties. The new method should be helpful to accelerate DEM simulations for systems associated with aggregates or agglomerates.
  \end{abstract}

  \begin{keyword}
  Discrete Element Method,
  Reduced stiffness,
  Microspheres,
  Cohesive particles,
  Rolling resistance,
  Packing structure

  \end{keyword}

\end{frontmatter}

%\linenumbers

\section{Introduction}
In multiphase and granular flows, discrete element method (DEM) has been widely used to model particle-particle interaction and accurately predict the motion of individual particles \citep{CundallGEO1979, TsujiPT1993, ZhuCES2008, Marshall2014, SundaresanAR2018, Xiao}. For soft-sphere DEM, Young's modulus of particles used in the simulation is usually much smaller than its real value. Therefore, it is reasonable to select a much larger time step to resolve inter-particle collisions, which considerably reduces the computation cost \citep{TsujiPT1993}. For systems with non-adhesive particles, the stiffness can be reduced by several orders without altering the simulation results. For instance, in fluidization systems, the flow patterns, both the shape and size distributions of bubbles, are reported to be insensitive to the particle stiffness \citep{MorenoCES2007, GuPT2016}. For a collision between two particles, the post-collisional velocity is also mainly determined by the damping coefficient rather than the particle stiffness \citep{MarshallJCP2009}.

However, for fine particles with van der Waals adhesion or wet particles with cohesion, a reduction of stiffness in DEM models can substantially change the simulation results \citep{GuPT2016}. Intuitively, with a smaller stiffness, the particles in contact tend to have a larger deformation along the direction of compression and an enlarged area of the contact region, which leads to an overestimation of the adhesive effect \citep{KobayashiPT2013, LiuCES2016}. To counterbalance the deviation that arises from the reduced stiffness, a modification of the adhesive force is often needed. \cite{KobayashiPT2013} simply regarded the adhesive force as a constant external force and then reduced it to recover the original value of the critical sticking velocity. Similar ideas were adopted by \cite{GuPT2016} and by \cite{HaervigPT2017}, who modified the van der Waals force between particles to conserve the cohesive energy during a quasi-static two-particle collision, and by \cite{WashinoPT2018} who derived a series of generic scaling to modify external attractive forces. In these previous studies, the adhesive force model has been modified based on the simple case of binary collision, thus is suitable only for the collision-dominated process, like fluidization process with a velocity much higher than the minimum fluidization value \citep{KobayashiPT2013, GuPT2016} or powder flow in a mixer with high rotating rate \citep{WashinoPT2018}.

An important but missing component in current adhesive DEM with reduced stiffness is its applicability to the problem associated with particle aggregates. These systems are actually quite ubiquitous in both industry and nature, such as coagulation of particles in interstellar space and protoplanetary disks \citep{ChokshiAJ1993, DominikAJ1997}, formation of dust cake during capture of aerosol particles \citep{LiJAS2007, ChenPRE2016, Wei}, packing of adhesive particles around or below the minimum fluidization velocity \citep{ValverdePRL2004, HakimPT2005, LiuSM2015, Shen}, etc. In these situations, it is of central importance to correctly predict both dynamic evolution and static structures of aggregates. For instance, it has been reported that coagulation between aggregates and particles dominates the growth from submicron-sized grains to kilometer-size planetesimals \citep{DominikAJ1997}, and the structures of dendrites (chainlike agglomerates) on fiber surfaces have a remarkable influence on capture efficiency of aerosol particle \citep{LiJAS2007, PayatakesCES1980}. Previous work has shown that packing of adhesive particles can be mechanically stable with packing fraction as low as 0.15 and coordination number close to 2 \citep{LiuSM2015, LiuPT2016, LiuSM2017}. Generally, a loose aggregate can be stable due to: (1) attractive forces that prevent the detaching of two contact particles (known as necking effect); (2) a rolling resistance and a sliding friction, arising from the asymmetrical distribution of the stress and surface roughness in the contact area, respectively. These resistances prevent the particle from rolling or sliding over its neighboring particles and should be properly calculated when reduced stiffness is used.

In this paper, we demonstrate how to properly consider rolling and friction resistances in the framework of adhesive DEM based on {\color{black}{Johnson-Kendall-Roberts (JKR) contact theory \citep{JKR}}} with reduced stiffness. It allows us to reproduce essentially the same packing structures as those calculated with real particle properties. Before showing a mass of packing structures in Section 4, we first briefly introduce the adhesive DEM based on JRK contact theory and rigorously derive a simple scaling law for use of reduced stiffness in Section \ref{sec_JKR}. In Section \ref{sec_inversion}, we propose an inversion procedure, by which the parameters in DEM can be set according to a prescribed particle-wall collision result.

\section{Modeling framework}
\label{sec_JKR}
\subsection{Adhesive DEM based on JKR contact theory}
The discrete element method is a framework that solves Newton's second law of each particle. The particles are regarded as soft bodies and the forces and torques between contact particles are resolved. In the JKR-based models proposed in our previous work \citep{LiJAS2007, MarshallJCP2009, LiPECS2011}, the normal force $F^N$, the sliding friction $F^S$, the twisting torque $M^T$, and the rolling torque $M^R$ acting on particle $i$ from its neighboring particle $j$ can be expressed as
% equation  for normal force and
  \begin{subequations}
    \begin{align}
  F_{ij}^N & \!=\!-4F_C\left(\hat{a}^3_{ij} \!-\! \hat{a}_{ij}^{3/2} \right) \!-\! \eta_N \bm{v}_{ij}\cdot \bm{n}_{ij}, \label{eqdem a} \\
  F_{ij}^S &\!=\! -\mathrm{min}\left[ k_T\int_{t_0}^t \bm{v}_{ij}(\tau)\cdot \bm{\xi}_S \mathrm{d}\tau \!+\!\eta_T\bm{v}_{ij}\cdot \bm{\xi}_S,\              F_{ij,crit}^S \right],        \label{eqdem b} \\
   M_{ij}^T &\!=\! -\mathrm{min}\left[ \frac{k_Ta^2}{2}\int_{t_0}^t \bm{\Omega}_{ij}^T(\tau)\cdot \bm{n}_{ij} \mathrm{d}\tau \!+\! \frac{k_Ta^2}{2}\bm{\Omega}_{ij}^T\cdot \bm{n}_{ij},\ M_{ij,crit}^T \right], \label{eqdem c} \\
  M_{ij}^R &\!=\! -\mathrm{min}\left[ 4F_C\hat{a}_{ij}^{3/2}\int_{t_0}^t \bm{v}_{ij}^L(\tau)\cdot \bm{t}_R \mathrm{d}\tau \!+\!\eta_R \bm{v}_{ij}^L\cdot \bm{t}_R,\ M_{ij,crit}^R \right]. \label{eqdem d}
\end{align}
\end{subequations}
The first term in the right-hand side of the normal force is derived from the JKR contact theory. It combines the effects of van der Waals attraction and elastic deformation of contact particles \citep{JKR}. The scale of the first term is set by the critical pull-off force, $F_C = 3\pi R_{ij}\gamma$, where $\gamma$ is the surface energy of the particle, $R_{ij} = (r_{p,i}^{-1} +r_{p,j}^{-1})^{-1}$ is the effective particle radius, $r_{p,i}$ is the radius of particle $i$. $\hat{a}_{ij}$ is calculated by normalizing the radius of the contact area $a_{ij}$ with its value at the zero-load equilibrium state $a_{ij,0}$, given as $a_{ij,0} = (9\pi\gamma R_{ij}^2/E_{ij})^{1/3}$ \citep{MarshallJCP2009}, $E_{ij}$ is the effective elastic modulus. The second term of Eq. (\ref{eqdem a}) is the viscoelastic dissipation, which is in propotion to the rate of deformation $\bm{v}_{ij}\cdot\bm{n}_{ij}$, {\color{black}{with $\bm{n}_{ij}$ being the unit vector pointing from the centroid of particle $i$ to that of particle $j$} and $\bm{v}_{ij} = \bm{v}_i - \bm{v}_j$ the relative particle velocity}. The normal dissipation coefficient $\eta_N = \alpha \sqrt{m^*k_N}$ is described in \citep{MarshallJCP2009, TsujiPT1992} with the coefficient $\alpha$ related to the coefficient of restitution $e$. {\color{black}$m^* = (m_i^{-1}+m_j^{-1})^{-1}$ is the effective mass of the two contacting particles, where $m_i$ is the mass of particle $i$.} The normal elastic stiffness $k_N$ is expressed as $k_N = \frac{4}{3}E_{ij}a_{ij}$ and the tangential stiffness $k_T$ is given as $k_T = 8G_{ij} a_{ij}$. The effective elastic and shear moduli $E_{ij}$ and $G_{ij}$ are both functions of particle’s Young’s modulus $E_i$ and Poisson ratio $\sigma_i$:
\begin{equation}
  \label{eqelastic}
  \frac{1}{E_{ij}} = \frac{1-\sigma_i^2}{E_i} + \frac{1-\sigma_j^2}{E_j},\quad \frac{1}{G_{ij}} = \frac{2-\sigma_i}{G_i} + \frac{2-\sigma_j}{G_j},
\end{equation}
where $G_i = E_i/2(1+\sigma_i)$ is the particle’s shear modulus. The sliding friction, the twisting torque, and the rolling torque (Eq. \ref{eqdem b} - \ref{eqdem d}) are all calculated using spring-dashpot-slider models, where $\bm{v}_{ij}\cdot\bm{\xi}_S$, $\bm{\Omega}_{ij}^T$, and $\bm{v}_{ij}^L$ are the relative sliding, twisting, and rolling velocities, {\color{black}{and $\bm{\xi}_S$ and $\bm{t}_R$ are the unit vectors in the directions of sliding and rolling, respectively}}. When these resistances reach their critical limits, $F_{ij,crit}^S$, $M_{ij,crit}^T$ or $M_{ij,crit}^R$, a particle will irreversibly slide, twist or roll relative to its neighboring particle. The critical limits are expressed as \citep{MarshallJCP2009}:
\begin{subequations}
\begin{align}
  F_{ij,crit}^S &= \mu F_C \left|4\left(\hat{a}_{ij}^3 - \hat{a}^{3/2}_{ij}\right) + 2 \right|, \label{eqcrit_a} \\
  M_{ij,crit}^T &= \frac{3\pi a_{ij} F_{ij,crit}^S} {16}, \label{eqcrit_b}\\
  M_{ij,crit}^R &= 4F_C\hat{a}_{ij}^{3/2} \theta_{crit}R_{ij}. \label{eqcrit_c}
\end{align}
\end{subequations}
Here $\mu$ is the friction coefficient and $\theta_{crit}$ is the critical rolling angle. One can set their values according to experimental measurements \citep{SumerJAST2008}.

\subsection{Accelerating adhesive DEM using reduced stiffness}
\begin{figure}
  \centering
  \includegraphics[width = 8 cm]{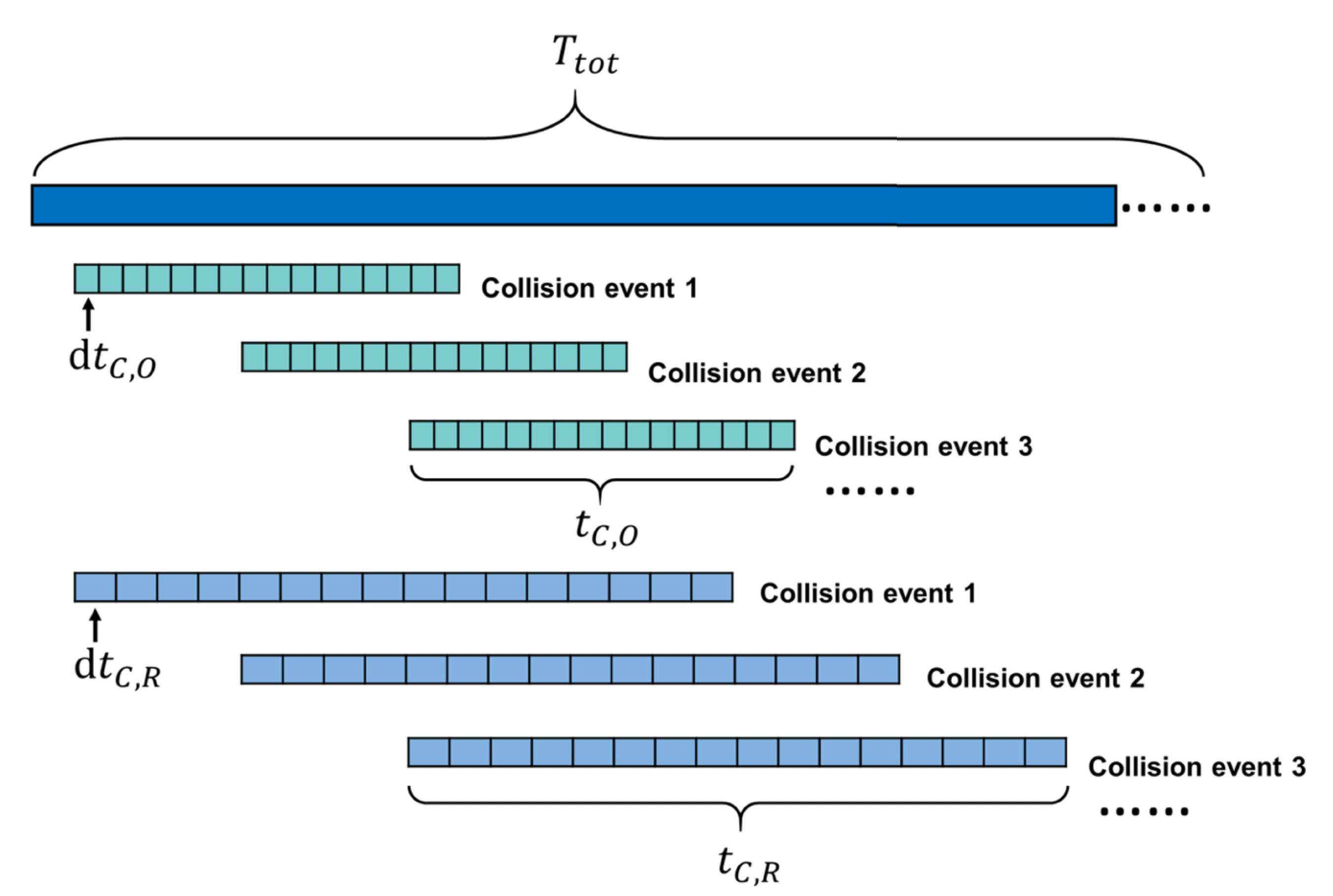}
  \caption{A graphical representation of accelerating DEM with reduced stiffness. The top dark blue bar indicates the entire simulated process, which has a time span $T_{tot}$. {\color{black}{The green bars stand for collision events calculated using the original stiffness of the particles, which have a typical timescale $t_{C,O}$, and the light blue bars are collision events calculated using the reduced stiffness of the particles, which have a timescale $t_{C,R}$}}. Each collision event is resolved by the time step $\mathrm{d}t_{C,O}$ or $\mathrm{d}t_{C,R}$, indicated by the discretized grids.}
  \label{fig_dt}
\end{figure}

{\color{black}{The typical collision time $t_C$, which is defined as the time associated with the elastic response during the collision between two particles, can be generally estimated as $t_C = r_p(\rho_p^2/E^2U)^{1/5}$ \citep{LiJAS2007}.}} To resolve the collision, one should use a time step $\mathrm{d}t_C = f_C t_C$ with $f_C$ much less than unity. Acceleration of the simulation can be achieved by choosing a reduced Young's modulus $E_R$ that satisfies the condition $E_R \ll E_O$ (hereafter, we use subscripts $O$ to indicate original particle properties and $R$ to indicate reduced properties). It allows one to use a larger time step to resolve the collision event. Such speedup of DEM is of prime importance when the simulated system contains numerous collision events. A graphical representation of this idea is displayed in Fig. \ref{fig_dt} The time span $T_{tot}$ is usually set by macroscopic parameters, such as the total mass loading of deposited particles in filtration/deposition systems or the total amount of gas in fluidization system, and thus is independent of the particle stiffness. {\color{black}{When the original Young's modulus $E_O$ is used in the simulation, the collision events (indicated by green bars) take place over the typical collision time $t_{C,O}$. In contrast, if a reduced Young's modulus $E_R$ is assigned to the particles, the collision events (indicated by light blue bars) will have a much larger timescale $t_{C,R}$. As a result, a larger time step $\mathrm{d}t_{C,R} (\gg \mathrm{d}t_{C,O})$ can be used to resolve the collision events and the total number of iterations decreases. Note that the collision events start at the same time when a reduced stiffness is used. Since the collision time is usually several orders of magnitude smaller than the typical particle transport time. It is reasonable to assume that the extension of the collision time due to the reduction does not apparently affect the start time of subsequent collision events.}}

%figure 1

\begin{figure}[h]
  \centering
  \includegraphics[width = 7.5 cm]{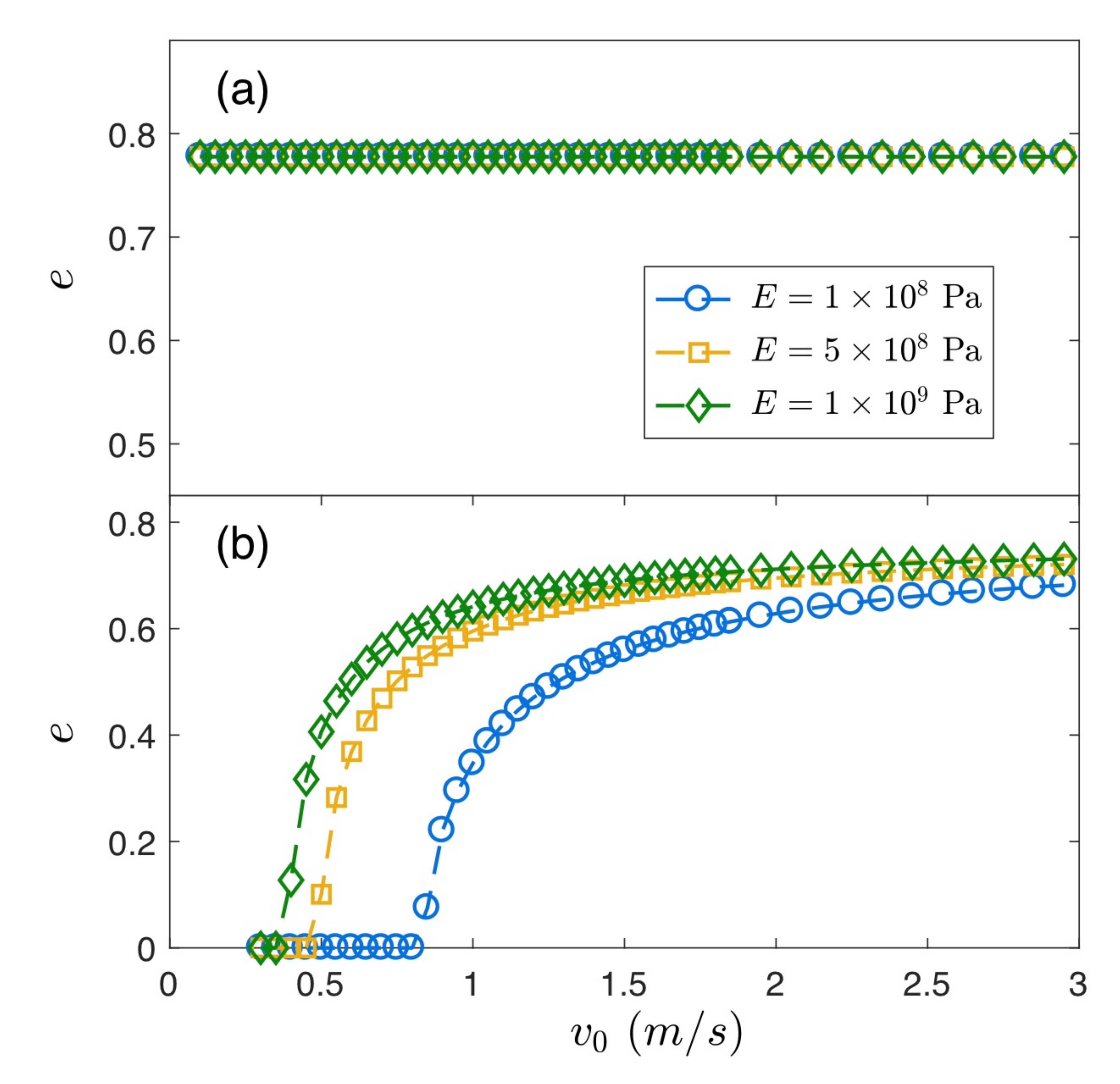}
  \caption{The coefficient of restitution $e$ as a function of impact velocity $v_0$ for particles with Young’s modulus $E = 10^8\ \mathrm{Pa}$ (circles), $5 \times 10^8\ \mathrm{Pa}$ (squares), and $10^9\ \mathrm{Pa}$ (diamonds). Panel (a) is for non-adhesive particle simulated by Hertz model, (b) is for adhesive particle described by JKR model. }
  \label{fig_ev1}
\end{figure}

%\subsection{Comparison of adhesive and non-adhesive Particle-wall collision}
Such approach for speedup of DEM has been tested and found widespread uses in the simulation of non-adhesive particles {\color{black}{\citep{MorenoCES2007, LommenPART2014, GuPT2016}}}. However, it cannot be directly applied to predict the behavior of adhesive small particles. The reason can be clearly illustrated through a simple case of the particle-wall collision. We consider a particle with radius $r$ and impact velocity $\mathrm{d}x/\mathrm{d}t = -v_0$. The state of the particle can be described using the equation of the overlap $\delta(t)$. {\color{black}{According to Newton's second law, the temporal evolution of $\delta(t)$ is given by $m \frac{\rm{d}^2\delta}{\rm{d}t^2} = F$. For non-adhesive particles, the force $F$ is calculated by the Hertzian model $-\frac{4E\sqrt{r}}{3}\delta^{3/2}$ together with damping force $-\eta_N\frac{\mathrm{d}\delta}{\mathrm{d}t}$ and the equation of $\delta$ takes the form \citep{TsujiPT1992, MarshallJCP2009}:}}
\begin{equation}
  \label{eq_herts}
  \frac{\mathrm{d}^2\delta}{\mathrm{d}t^2} + \frac{\eta_N}{m}\frac{\mathrm{d}\delta}{\mathrm{d}t} + \frac{4E\sqrt{r}}{3m}\delta^{3/2} = 0.
\end{equation}
In Hertzian model, the radius of contact region is calculated as $a = \sqrt{r\delta}$. {\color{black}{For adhesive particles, instead of using the Hertzian model, we use the JKR model (Eq. (\ref{eqdem a})) to calculate the forces. The equation of $\delta$ now becomes}}
\begin{equation}
  \label{eqode}
\frac{\mathrm{d}^2\delta}{\mathrm{d}t^2} + \frac{\eta_N}{m}\frac{\mathrm{d}\delta}{\mathrm{d}t} + \frac{4F_C}{m}\left( \hat{a}^3(\delta) - \hat{a}^{3/2} (\delta)\right) = 0.
\end{equation}
Here, $\hat{a}$ is related to the overlap $\delta$ through \citep{MarshallJCP2009}
\begin{equation}
  \label{eqavsd}
\frac{\delta}{\delta_C} = 6^{1/3}\left[2(\frac{a}{a_0})^2 - \frac{4}{3}(\frac{a}{a_0})^{1/2} \right].
\end{equation}
The critical overlap $\delta_C$ is given by $\delta_C = a^2_0/(2(6)^{1/3}r)$. The result of a collision is described using the coefficient of restitution $e$, defined as the ratio of the post-collision velocity of the particle to its velocity before the collision. To show the effect of reduced particle stiffness, three different values of Young’s modulus, $E = 10^9$ Pa, $5\times 10^8$ Pa, and $10^8$ Pa are used. As shown in Fig. \ref{fig_ev1}, the restitution coefficient $e$ for collisions between the non-adhesive particle and the wall is independent of the particle’s Young’s modulus. Whereas, for the adhesive particle, the restitution coefficient, especially at a low impact velocity, significantly decreases when Young’s modulus is reduced. And a remarkable increase of the critical sticking velocity $v_C$, {\color{black}{defined as the maximum impact velocity at which a particle hitting the surface will stick (i.e., $e=0$),}} is also observed. To address this issue, it has recently been suggested that a reduced surface energy should be used to balance the nonphysical effect caused by reduced stiffness so that the outcome of the collision will remain the same \citep{HaervigPT2017, WashinoPT2018}.

Here, we derive a scaling law for choosing the reduced surface energy in a rigorous way based on the non-dimensional equation for particle collision. We start from the non-dimensional form of the equation describing the collision between an adhesive particle and a wall (Eq. (\ref{eqode}))
\begin{equation}
  \label{eqode2}
  \frac{\mathrm{d}^2\hat{\delta}}{\mathrm{d}\hat{t}^2} + A \hat{a}^{1/2}\frac{\mathrm{d}\hat{\delta}}{\mathrm{d}\hat{t}} + Bg(\hat{\delta}) = 0.
\end{equation}
The overlap is normalized by the critical overlap $\delta_C$, and the time is scaled using $T_0 = \delta_C/v_0$. The coefficients $A$ and $B$ are functions of particle properties and the initial velocity $v_0$:
\begin{subequations}
  \begin{align}
    A &= 2.515\alpha \left( \frac{E}{\rho v_0^2} \right)^{-\frac{1}{3}}\left(\frac{\gamma}{\rho v_0^2 r} \right)^{\frac{5}{6}}, \label{eqA} \\
    B &= \frac{3.633}{\alpha^2}A^2. \label{eqB}
  \end{align}
\end{subequations}
$\hat{a}$ in Eq. (\ref{eqode2}) can be calculated inversly through Eq. (\ref{eqavsd}) and $g(\hat{\delta})$ is expressed as $g(\hat{\delta}) = \hat{a}^3 - \hat{a}^{3/2}$. Eq. (\ref{eqode2}) can be solved given the initial conditions $\hat{\delta}(0) = 0$ and $\frac{\mathrm{d}\hat{\delta}}{\mathrm{d}\hat{t}}(0) = 1$, and the jump-on/jump-off criterion: the contact between the particle and wall is built up when $\hat{\delta} > 0$ and is broken up when $\hat{\delta} < -1$. The result of a collision is determined only by the damping coefficient $\alpha$ and a grouped parameter $A^*$, which is defined as
\begin{equation}
  \label{eqAstar}
  A^* = \mathcal{H}(E,\gamma, \rho, v_0) = \left(\frac{E}{\rho v_0^2} \right)^{-\frac{1}{3}} \left( \frac{\gamma}{\rho v_0^2 r} \right)^{\frac{5}{6}} \equiv (El)^{-\frac{1}{3}}(Ad)^{\frac{5}{6}}.
\end{equation}
We have removed all other coefficients in the governing equation and in the initial conditions through scaling. The first parameter in Eq. (\ref{eqAstar}), {\color{black}{$El=\frac{E}{\rho v_0^2}$}}, is called elasticity parameter, which can be regarded as the ratio of elastic force to the particle inertia \citep{LiJAS2007}. The second parameter is the adhesion parameter, {\color{black}{$Ad=\frac{\gamma}{\rho v_0^2 r}$}}, defined as the ratio of the adhesive energy and the particle kinetic energy \citep{LiJAS2007}. {\color{black}{The adhesion parameter $Ad$ has been successfully used to estimate the critical sticking velocity of two colliding particles \citep{ChenPT2015} and predict the packing structure of adhesive particles \cite{LiuSM2015}. For monodispersed systems, $Ad$ is usually defined based on the reduced radius $R = r/2$ \citep{LiuSM2015}. In contrast, for polydisperse systems or systems where particles and walls coexist, a mean particle radius $<r>$ is preferred to calculate $Ad$.}}

\begin{figure}[h]
  \centering
  \includegraphics[width = 9cm]{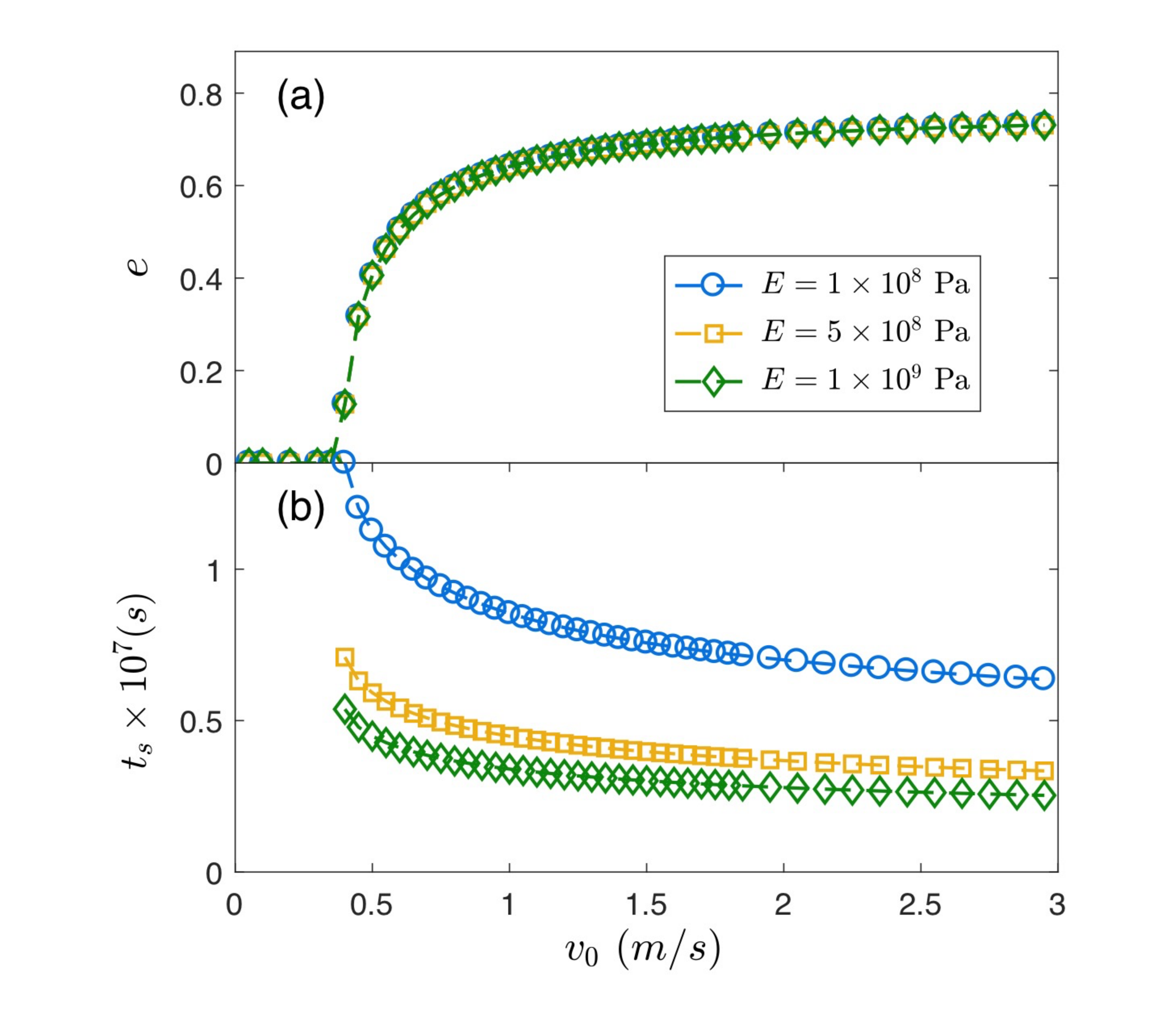}
  \caption{(a) The coefficient of restitution $e$ as a function of impact velocity $v_0$ for particles with Young’s modulus $E = 10^8$ Pa (circles), $5\times10^8$ Pa (squares), and $10^9$ Pa (diamonds). The surface energy is modified according to Eq. (\ref{eqgamr}). (b) Corresponding time interval $t_S$ between the moment of contact formation and the moment of separation.}
  \label{fig_ev2}
\end{figure}

When a reduced particle Young’s modulus $E_R$ is used, the surface energy should be modified to keep $A^*$ constant. Thus, the reduced surface energy is calculated as:
\begin{equation}
  \label{eqgamr}
  \gamma_R = \left(\frac{E_R}{E_O} \right)^{\frac{2}{5}}\gamma_O \equiv \chi^{\frac{2}{5}}\gamma_O,
\end{equation}
where $\chi = E_R/E_O$ is the reduced ratio. We recalculate the particle-wall impact case in Fig. \ref{fig_ev1} (b) with surface energy modified according to Eq. (\ref{eqgamr}). The result in Fig. \ref{fig_ev2} (a) demonstrates that the scaling yields $e-v$ curves identical to those calculated with the original parameter. We also display the physical time of collisions $t_s$, which is defined as the time interval between the moment of contact formation and separation, as a function of impact velocity. For a given velocity, the collision takes place over a much longer time when reduced Young’s modulus is used. The simple scaling in Eq. (\ref{eqgamr}) has the same form with the one derived from the concept of energy conservation during quasi-static separation of two contact particles \citep{HaervigPT2017}. The derivation here is based on the dimensionless equation of the motion for true collision cases thus is believed to be more rigorous.

\subsection{Modified models for rolling and sliding resistances}
A proper description of adhesive rolling and sliding resistances is of significance to predict the formation of agglomerates and the structure of particle deposits. For adhesive microparticles, rolling is generally the preferred deformation mode, which gives rises to the rearrangement of packing structures \citep{DominikAJ1997, LiuPT2016, LiuSM2017}. To accurately simulate the rolling motion, the adhesive rolling model needs to be modified in the framework of JKR-based DEM with reduced stiffness. The same idea can be readily applied to modify the sliding resistance.

\begin{figure}[h]
  \centering
  \includegraphics[width = 9 cm]{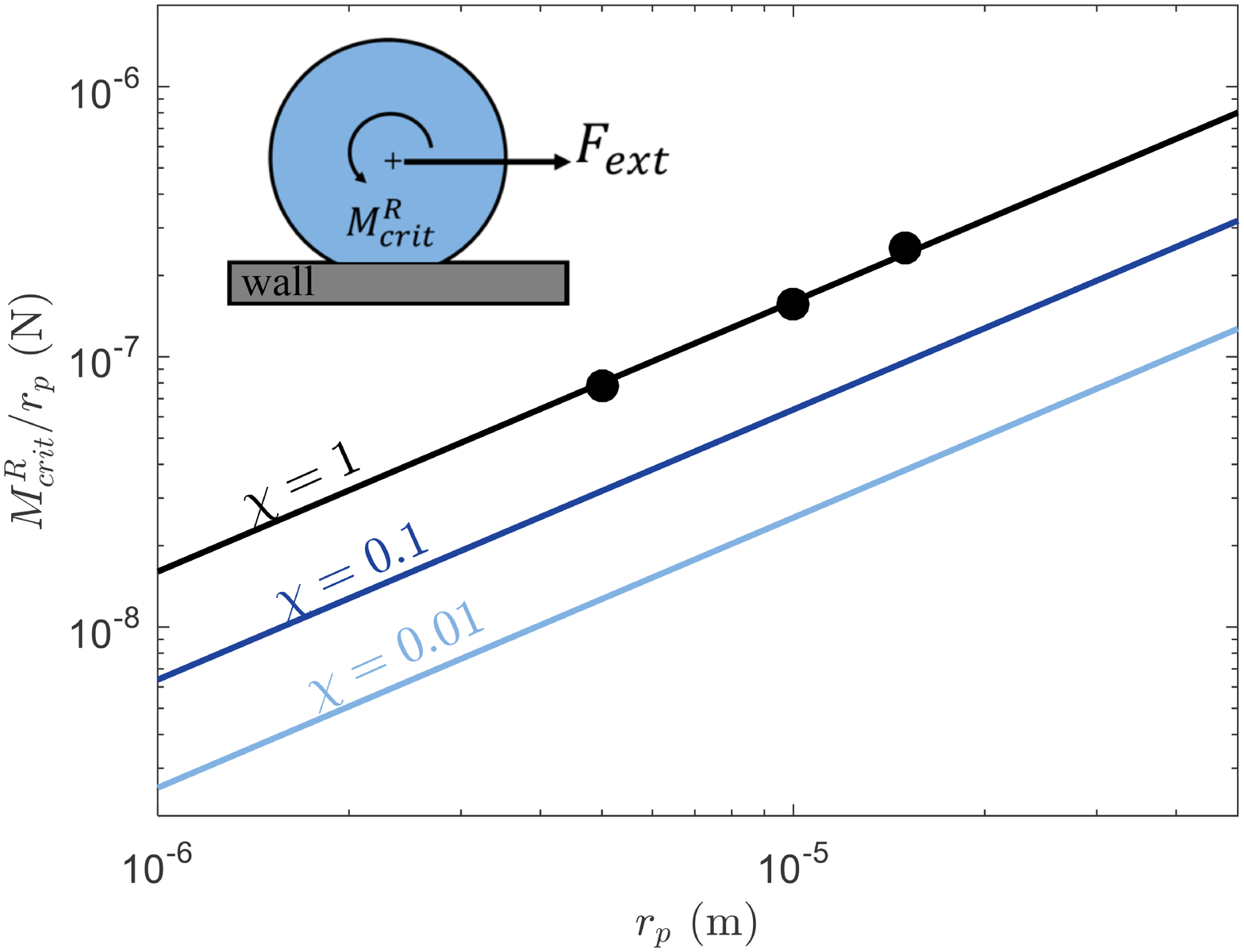}
  \caption{\label{fig_roll} The critical rolling force $M_{crit}^R/r_p$ as a function of particle size $r_p$ at different stiffness-reduced ratio $\chi = E_R/E_O$. The black circles are experimental results from \citep{SumerJAST2008}. The solid lines are calculations of Eq. (\ref{eqcrit_c}) using the surface energy $\gamma_R = \chi^{2/5} \gamma_O$ and the critical rolling angle $\theta_{crit} = \xi_{crit}/r_p= 0.0085$, with $\gamma_O$ and $\xi_{crit}$ the same as those measured in the experiments \cite{SumerJAST2008}. The inset shows the set-up of the measurements.}
\end{figure}

Assume a simple case where a particle is in normal equilibrium with a wall and an external force $F_{ext}$, which is parallel to the wall, is then applied on the center of the particle. If $F_{ext}$ is smaller than the critical value $M_{crit}^R/r_p$, the particle rolls over a small distance and reaches a mechanically stable state. If $F_{ext} > M_{crit}^R/r_p$, the particle will roll irreversibly. According to the experimental measurements of \cite{SumerJAST2008} using polystyrene microparticles, the critical rolling angle $\theta_{crit} = \xi_{crit}/r_p$ is nearly constant, $\theta_{crit} = 0.0085$. As displayed in Fig. \ref{fig_roll}, using the same parameters as in \citep{SumerJAST2008}, Eq. (\ref{eqcrit_c}) gives a good prediction of the particle size dependence of the critical rolling force. However, as displayed in Fig. \ref{fig_roll}, $M_{crit}^R/r_p$ is underestimated when a reduced particle stiffness $E_R = \chi E_O$ and the corresponding reduced surface energy $\gamma_R = \chi^{2/5}\gamma_O$ are used. The reason is that the critical rolling resistance in Eq. (\ref{eqcrit_c}) is proportional to the surface energy but is independent of particle stiffness. An easy and intuitive way to retain the original value of the critical rolling resistance is to use the real surface energy $\gamma_O$ to calculate the rolling resistance. {\color{black}{Substituting $F_C$ in Eq. (\ref{eqcrit_c}) with $F_C = 3\pi\gamma_OR_{ij}$, we have}}:
\begin{equation}
  \label{eqmodrolling}
  M_{crit}^R = \underbrace{12\pi \gamma_O R_{ij} \hat{a}_{ij}^{3/2}}_{{\rm rolling\ stiffness}\ k_r} \theta_{crit}R_{ij}.
\end{equation}

In some particular cases, where the friction coefficient is small enough (usually smaller than $0.05$) to yield $F_{crit}^S < M_{crit}^R/r_p$, irreversible sliding will be triggered before rolling \citep{LiuSM2017,LiuPT2016}. In such conditions, one should calculate the critical sliding forces $F_{crit}^S$ in Eq. (\ref{eqcrit_a}) using original value of particle properties, i.e.,
\begin{equation}
  \label{eqmodslide}
  F_{crit}^S = \mu \cdot (3\pi\gamma_OR_{ij}) \cdot \left|4\left(\hat{a}_{ij}^3 - \hat{a}_{ij}^{3/2} \right) + 2\right|,
\end{equation}
{\color{black}{where the critical pull-off force, $F_C$, in Eq. (\ref{eqcrit_a}) is again calulated using the original value of the particle surface energy $F_C = 3\pi\gamma_OR_{ij}$.}}

\section{An inversion procedure to set parameters in adhesive DEM}
\label{sec_inversion}
\begin{figure}
  \centering
  \includegraphics[width = 9 cm]{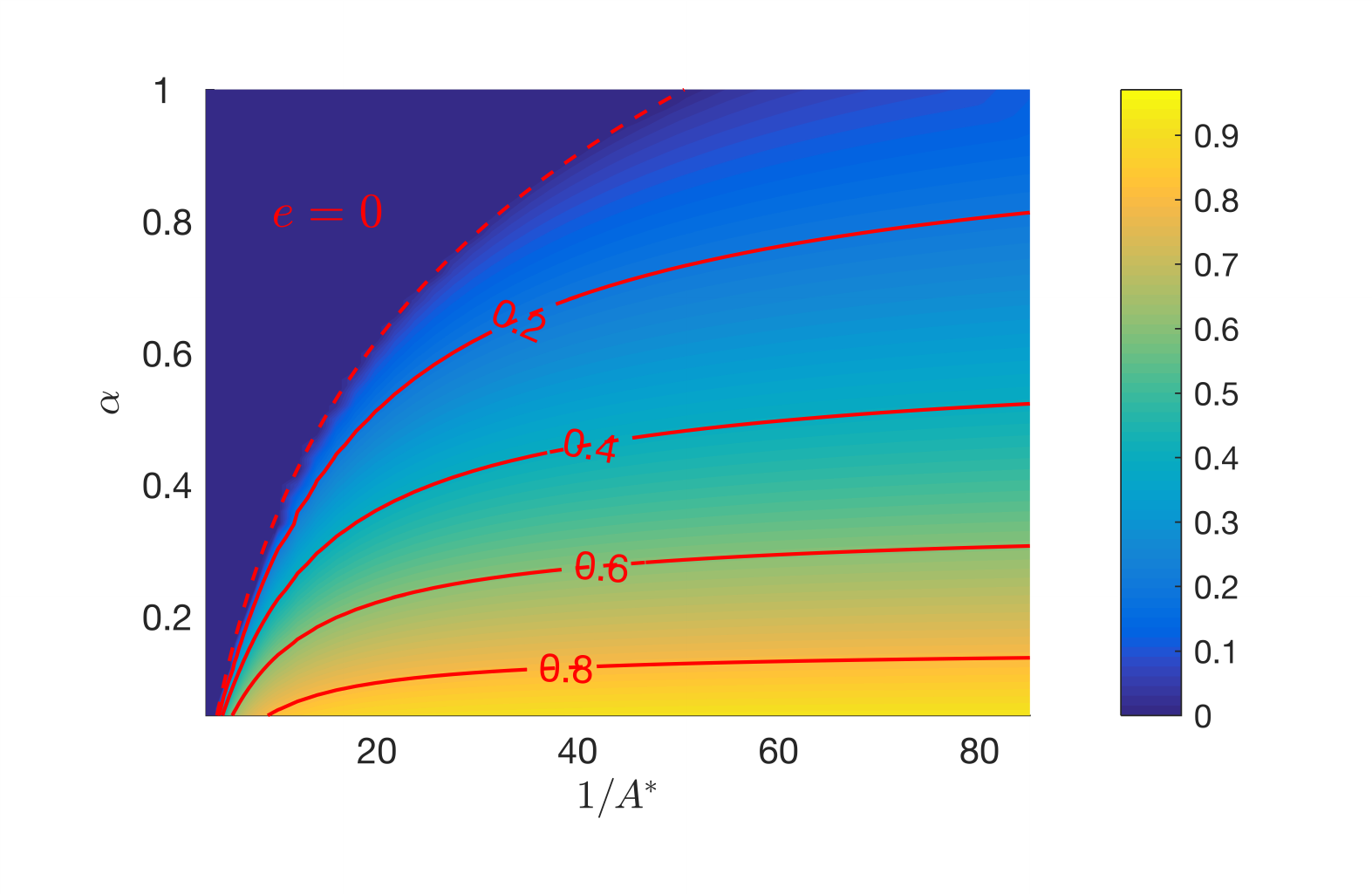}
  \caption{Coefficient of restitution $e$ as a function of damping coefficient $\alpha$ and the inverse of the parameter $A^*$. The value of $e$ is indicated by the color scale with red contour lines. The dashed red line separates the sticking region ($e = 0$) and the rebound region ($e>0$).}
  \label{fig_e}
\end{figure}

\begin{figure}
  \centering
  \includegraphics[width = 8.5 cm]{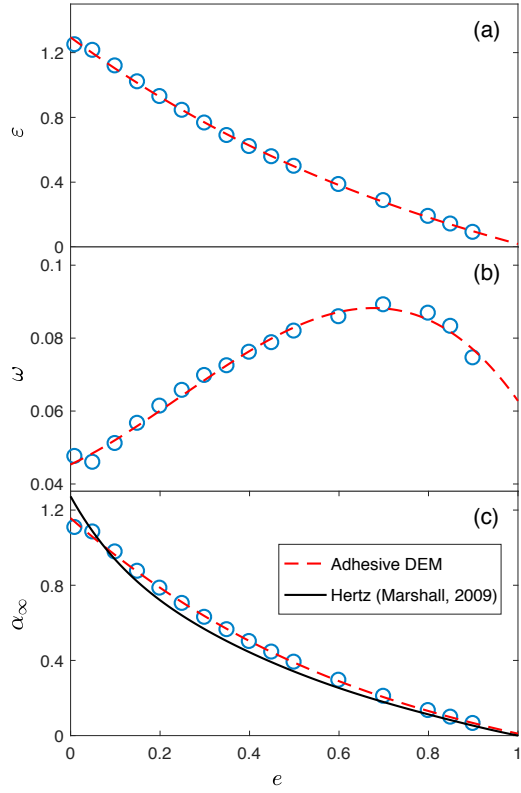}
  \caption{Fitting parameters $\varepsilon$, $\omega$ and $\alpha_{\infty}$ in Eq. (\ref{eqfitalpha}) as functions of the restitution coefficient $e$ (data points). Dashed lines are the three-order polynomial fittings of Eq. (\ref{eqfitall}). Solid line in (c) is the relationship between $\alpha$ and $e$ in \citep{MarshallJCP2009}, which is derived for non-adhesive particles based on Hertz model.}
  \label{fig_alpha}
\end{figure}

A principle for setting parameters in the framework of fast adhesive DEM with reduced stiffness is proposed in this section. {\color{black}{Based on Eq. (\ref{eqA}), the parameters to be determined include the damping coefficient $\alpha$, a reduced particle Young's modulus $E_R$ and a reduced surface energy $\gamma$. Ohter parameters in Eq. (\ref{eqA}) can be easily determined from direct measurement (particle density $\rho$ and radius $r$) or is regarded as an input parameter for simulations (initial velocity $v_0$). In Eq. (7), $\alpha$ and $1/A^*$ are the only parameters that will affect the result (i.e., the coefficient of restitution $e$.)}} The contour plot in Fig. \ref{fig_e} shows the value of restitution coefficient $e$ as a function of damping coefficient $\alpha$ and $1/A^*$. We use $1/A^*$ instead of $A^*$ because $1/A^*$ scales as $1/A^* \sim v_0$ and the initial collision velocity $v_0$ is usually a well-controlled parameter in experiments. Several interesting features can be observed: (1) there is a sticking region ($e = 0$) when both $\alpha$ and $A^*$ are large values; (2) with a large value of $1/A^*$, $e$ has a weak dependence on $1/A^*$ and is mainly determined by the dissipation coefficient $\alpha$. For instance, at $1/A^* > 40$, the contour lines with $e = 0.6$ and $e = 0.8$ are nearly parallel to the abscissa axis. {\color{black}{For any given $e$, the coutor lines approximately follow an exponential form. Based on this observation, we assume an exponential relation between $\alpha$ and $1/A^*$ with the fitting parameters $\varepsilon$, $\omega$ and $\alpha_{\infty}$ determined by $e$.}}
\begin{equation}
  \label{eqfitalpha}
  \alpha = \alpha_{\infty} - \varepsilon \mathrm{exp}\left(-\frac{\omega}{A^*} \right).
\end{equation}
{\color{black}{We fit the coutor lines in Fig. \ref{fig_e} using Eq. (\ref{eqfitalpha}) with $e$ varing from $\sim 0$ to $0.9$. And the fitting parameters $\varepsilon$, $\omega$ and $\alpha_{\infty}$ are all inversely calculated from $e$ through three-order polynomial fittings (as shown in Fig. \ref{fig_alpha}).}}
\begin{subequations}
  \label{eqfitall}
\begin{align}
  \varepsilon(e)       &= -0.2302e^3 + 0.9806e^2 - 2.026e   + 1.294,   \label{eqfitalla} \\
  \omega(e)            &= -0.1504e^3 + 0.110e^2  + 0.05783e + 0.04534, \label{eqfitallb} \\
  \alpha_{\infty}(e)  &= -0.3325e^3 + 1.279 e^2 - 2.094 e  + 1.157.   \label{eqfitallc}
\end{align}
\end{subequations}

Based on Eqs. (\ref{eqfitalpha}) and (\ref{eqfitall}), an inversion procedure to determine the value of $\alpha$ and $A^*$ in DEM from the experimental data is proposed as:

\begin{enumerate}[(a)]
\item Use Eq. (\ref{eqfitallc}) to determine $\alpha (= \alpha_{\infty})$ according to the value of $e$ at {\color{black}{high-velocity}} state ($A^{*-1} \to \infty $).
\item Pick another typical point on $e-v$ curve, $(v_t, e_t)$, and calculate corresponding $\varepsilon_t$, $\omega_t$ and $\alpha_{\infty,t}$ through Eq. (\ref{eqfitall}).
\item Using the values of $\alpha$ obtained from step (a) and the parameters $\varepsilon_t$, $\omega_t$ and $\alpha_{\infty,t}$ from (b), {\color{black}{calculate $A^*_t(\alpha; \varepsilon_t, \omega_t, \alpha_{\infty,t})$ inversely from Eq. (\ref{eqfitalpha}): $ A^*_t = -\omega_t \ln^{-1}(\frac{\alpha_{\infty,t} - \alpha}{\varepsilon_t})$.}}
\item Choose pseudo Young’s modulus $E_R$ and surface energy $\gamma_R$, which are usually much smaller than their original values $E_O$ and $\gamma_O$, and make sure $E_R^{-1/3}\gamma_R^{5/6} = A^*_t \rho^{1/2}v_tr^{5/6}$ (see Eq. (\ref{eqAstar})).
\end{enumerate}

 Step (a) is extended from the $e-\alpha$ relationship in non-adhesive collision cases, where $e$ is almost a constant that is determined by damping coefficient $\alpha$. Therefore, for non-adhesive particles, one can calculate damping coefficient $\alpha$ inversely from $e$. In Fig. \ref{fig_alpha} (c), we plot such a correlation: $\alpha = 1.2728 - 4.2783 e + 11.087 e^2  - 22.348 e^3 + 27.467 e^4 - 18.022 e^5 + 4.8218 e^6$, which is proposed by \citep{MarshallJCP2009}. When an adhesive particle collides with a wall, $e$ is zero if $v_0$ is smaller than the critical sticking velocity $v_c$. As $v_0$ increases, $e$ will first increase and then enter a plateau, corresponding to the region $\frac{\partial e}{\partial \alpha} \gg \frac{\partial e}{\partial (1/A^*)}$ in Fig. \ref{fig_e}. In the {\color{black}{high-velocity}} state, the amount of energy dissipated due to viscoelasticity is much larger than that of first-contact loss (i.e., necking effect). As a result, the function $e(\alpha, A^*)$ reduces to a single-parameter function $e_{\infty}(\alpha)$ and we relate $\alpha$ to $e_{\infty}$ through Eq. (\ref{eqfitallc}) (Step (a)). As shown in Fig. \ref{fig_alpha} (c), there is only a slight difference between the $\alpha - e$ curve calculated from adhesive DEM and that from Hertz model.

Recall that $A^* = \mathcal{H}(E,\gamma,\rho, v_0)$ is a function of particle properties and the velocity. One may expect to determine $A^*$ according to the real physical properties of the particle and further predict the restitution coefficient $e$. However, these parameters are usually not readily available. For example, the surface energy $\gamma$ is strongly affected by the surface roughness and the ambient humidity and is usually hard to determine. In addition, to accelerate the computation, a reduced Young’s modulus instead of its true value is often needed. From step (b) to (d), we suggest one to alternatively select another typical point on a prescribed $e-v$ curve $(v_t, e_t)$ that is outside the {\color{black}{high-velocity}} region and use Eq. (\ref{eqfitalpha}) to obtain the corresponding $A_t^*(e_t, \alpha)$ and to further get the value of $E_R^{-1/3}\gamma_{R}^{5/6}$ through $E_R^{-1/3}\gamma_{R}^{5/6} = A_t^*\rho^{1/2}v_tr^{5/6}$, which can reproduce the prescribed $e-v$ curve.

\begin{figure}
  \centering
  \includegraphics[width = 10 cm]{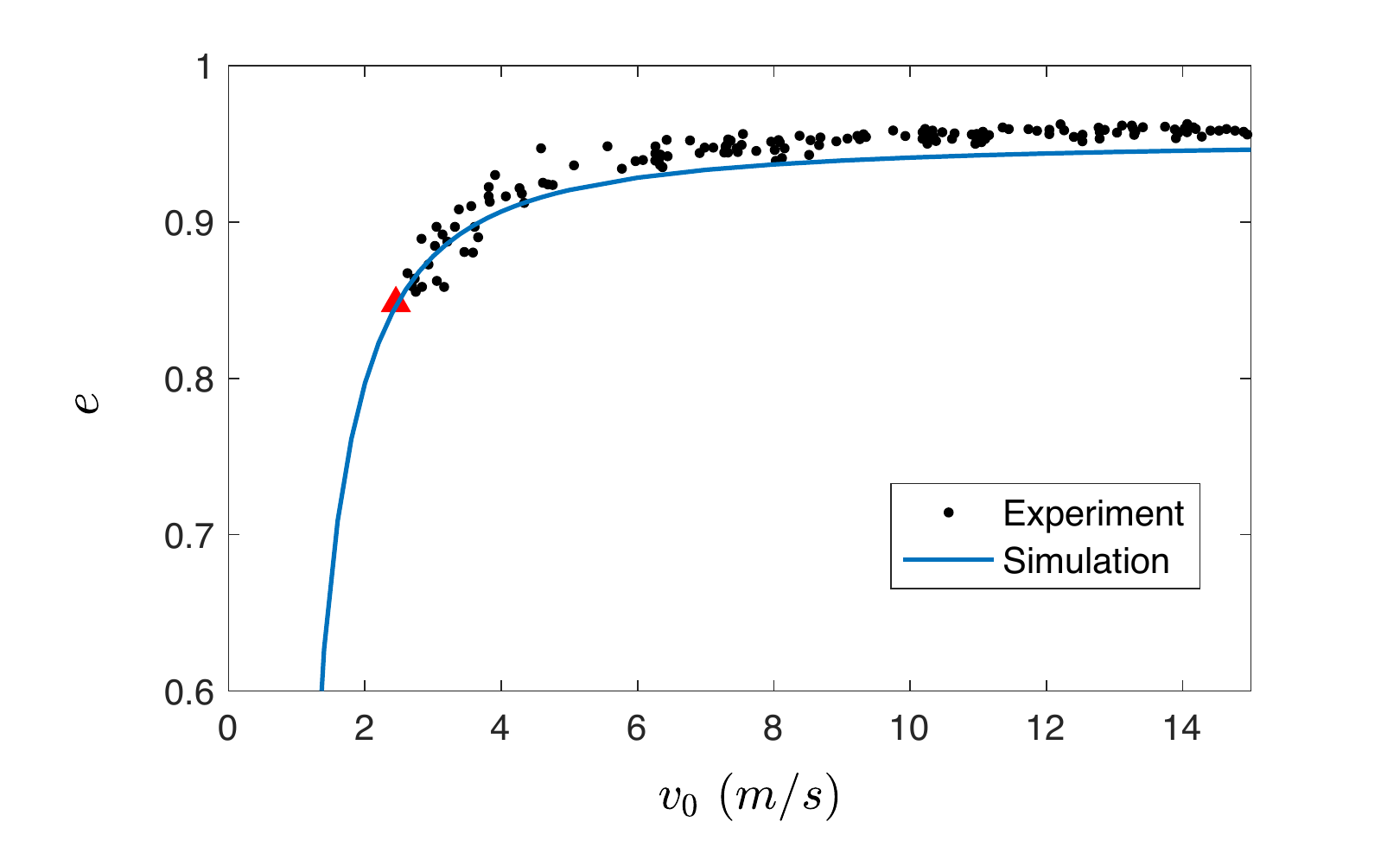}
  \caption{Comparison of the coefficient of restitution $e$ calculated by JKR-Based DEM to the experimental measurements from \citep{DahnekeJCIS1975}. Parameters used in DEM are determined through our inversion procedure. The red triangle stands for the typical point at $v_0 = 2.454$ and $e = 0.848$.}
  \label{fig_exp}
\end{figure}

In Fig. \ref{fig_exp} we present an example of the inversion procedure based on the experimental data of \citep{DahnekeJCIS1975}: (a) Use Eq. (\ref{eqfitallc}) and the coefficient of restitution in the {\color{black}{high-velocity}} region, $e = 0.96$, to obtain {\color{black}{$\alpha_{\infty} = 0.0321$}}; (b) Pick a typical point ($v_t, e_t$) on $e-v$ curve -- here we use the point ($2.454, 0.848$), indicated by the triangle in Fig. \ref{fig_exp} -- and then calculate the fitting parameters $\varepsilon_t$, $\omega_t$ and $\alpha_{\infty,t}$ at $e_t = 0.848$, then (c)  solve Eq. (\ref{eqfitalpha}) to obtain $A_t^* =0.11$. (d) Determine the value of $E^{-1/3}\gamma^{5/6}$ through: $E_R^{-1/3}\gamma_R^{5/6} = A^*_t\rho^{1/2}v_tr^{5/6} = 5.92\times 10^{-5}\ \mathrm{N^{1/2}m^{-1/6}}$. The value obtained in this way is quite close to the value calculated using physical properties of polystyrene particles (PSL): $E_O^{-1/3}\gamma_O^{5/6} = (3.8\ \mathrm{GPa})^{-1/3}(0.05\ \mathrm{Jm^{-2}})^{5/6} = 5.28\times 10^{-5}\ \mathrm{N^{1/2}m^{-1/6}}$ \citep{SumerJAST2008}. At last, pick a reduced Young’s modulus $E_R$ and calculate the modified $\gamma_R$. As shown in Fig. \ref{fig_exp}, the $e-v$ curve calculated from $E_R^{-1/3}\gamma_R^{5/6}$ well reproduces the experimental measurements \citep{DahnekeJCIS1975}. A large number of research has reported experimental results of $e-v$ curves \citep{WallAST1990, DunnAST1995, LiJAS1999, KimJAS2008, SoraceMRC2009}, and the proposed inversion procedure is easy to run to assist the selection of contact parameters before large-scale DEM simulations.

\section{Test on packing problem}
\begin{figure}
  \centering
  \includegraphics[width = 6 cm]{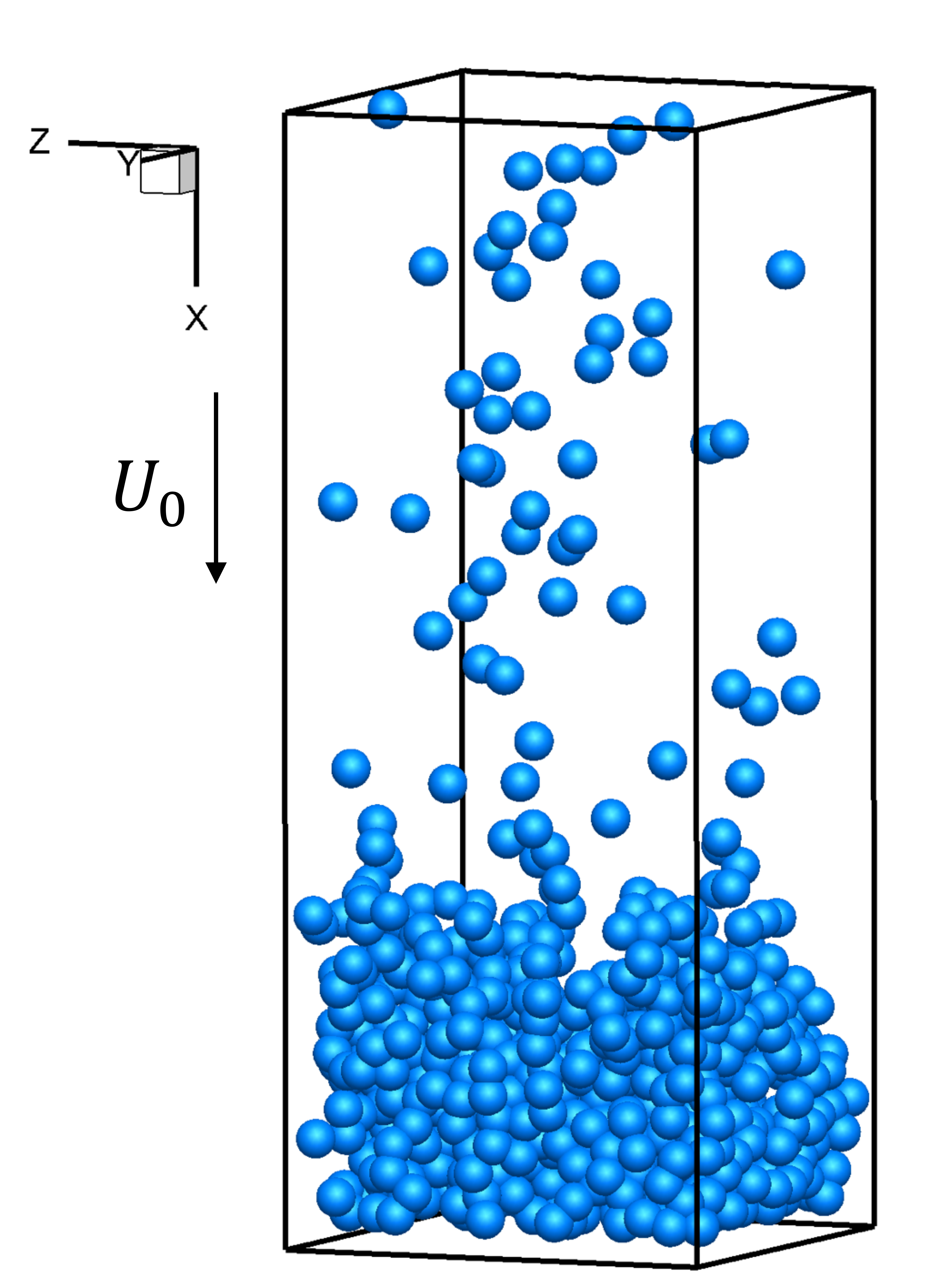}
  \caption{Schematic of simulation setup.}
  \label{fig_pk_setup}
\end{figure}

To check if the fast adhesive DEM can reproduce the results calculated using original particle properties in cases associated with aggregates. We run a large number of cases on the packing problem. As shown in Fig. \ref{fig_pk_setup}, we consider ballistic falling of $N(=2000)$ particles.  Particles have radius $r_p$ and initial velocity $\bm{U}_0 (=(U_0,0,0))$ and are randomly added into the computational domain from an inlet plane at height $L_x (=160r_p)$. Periodic boundary conditions are set along $y$ and $z$ directions with box length $L_y = L_z =28r_p$. The physical parameters used in our simulations are set according to the properties of polystyrene (PS) particle in \citep{SumerJAST2008}, which has the density $\rho = 1000\ \mathrm{kg/m^3}$, Young’s modulus $E_O = 3.8\times 10^9\ \mathrm{Pa}$, surface energy $\gamma_O = 0.05\ \mathrm{J/m^2}$, friction coefficient $\mu_f = 0.3$, and the critical rolling angle $\theta_{crit} = 0.0085$. A vacuum condition is assumed to filter out fluid effect. Gravity effect can be neglected since the Froude number, $Fr = U_0/(gL_x)^{1/2}$ of our system satisfies $Fr \gg 1$. This ballistic packing system has been widely used in both experimental \citep{BlumPRL2004, ParteliSR2014} and numerical \citep{YangPRE2000, YangGM2013, LiuSM2015, LiuPT2016, LiuAICHE2017, LiuSM2017} studies and has been proved to be useful to bridge the gap between the particle-level interactions and the macroscopic structure of aggregates \citep{ChenSM2016, Baule}.

To understand how to simulate the packing process in the framework of fast adhesive DEM with reduced stiffness, we set 4 series of computational experiments (listed in Table \ref{tab:param}): in the cases of series S, we use the original value of the elastic modulus $E_O$ and surface energy $\gamma_O$ and the results can be regarded as a benchmark case; in series A, reduced elastic modulus $E_R$ is used without modification of the surface energy; in series B, we use the same elastic modulus as those in A and modify the surface energy according to $\gamma_R = \chi^{2/5}\gamma_O$; series C is essentially the same as series B except that the rolling stiffness is calculated based on the original surface energy, i.e., $k_r = 12\pi \gamma_O R_{ij} \hat{a}_{ij}^{3/2}$. For each case, at least 10 final configurations are obtained to provide a meaningful average and standard deviation. According to the analysis in Section 2, the packing structure is essentially determined by three parameters: the damping coefficient $\alpha$, which is fixed here, the dimensionless adhesive parameter $Ad$, and the elasticity parameter $El$. To separately tune the value of $Ad$ at given $El$, we fixed the velocity $U_0$ and varied the particle size $r_p$ in our simulation.

\begin{table}
  \caption{\label{tab:param} Parameters used in DEM simulations of microparticle packings. The parameters used in the case S are the same as those in the experiments \citep{SumerJAST2008} and are regarded as original particle properties. In series A, three reduced particle moduli are used without modification of surface energy. In series B, surface energy is modified according to $\gamma_R = \chi^{2/5}\gamma_O$. In case C, we modified the surface energy in the same fashion as in case B and use the original surface energy $\gamma_O$ to calculate the rolling stiffness, i.e. $k_r = 12\pi \gamma_O R_{ij} \hat{a}_{ij}^{3/2}$}
\begin{tabular}{l*{6}{c}r}
\hline
Parameters     & $E\ \mathrm{(Pa)}$  &$\gamma\ \mathrm{(J/m^2)}$&  $k_r$                          &   $El$            & $Ad$         \\
\hline
S              & $3.8\times 10^9$    & 0.05                     & $12\pi \gamma_O R_{ij} \hat{a}_{ij}^{3/2}$ & $1.69\times10^6$  & $0.1 \sim 33$   \\
\hline
A-1            & $1.0\times 10^9$    &                          &                                 & $4.44\times10^5$  &              \\
A-2            & $5.0\times 10^8$    & 0.05                     & $12\pi \gamma_O R_{ij} \hat{a}_{ij}^{3/2}$ & $2.22\times10^5$  & $Ad_S$       \\
A-3            & $1.0\times 10^8$    &                          &                                 & $4.44\times10^4$  &              \\
\hline
B-1            & $1.0\times 10^9$    & 0.0293                   &                                 & $4.44\times10^5$  &               \\
B-2            & $5.0\times 10^8$    & 0.0222                   & $12\pi \gamma_R R_{ij} \hat{a}_{ij}^{3/2}$ & $2.22\times10^5$  & $Ad_S\cdot \chi^{2/5}$ \\
B-3            & $1.0\times 10^8$    &  0.0117                  &                                 & $4.44\times10^4$  &               \\
\hline
C-1            & $1.0\times 10^9$    &  0.0293                  &                                  & $4.44\times10^5$  &            \\
C-2            & $5.0\times 10^8$    &  0.0222                  &  $12\pi \gamma_O R_{ij} \hat{a}_{ij}^{3/2}$ & $2.22\times10^5$  & $Ad_S\cdot \chi^{2/5}$ \\
C-3            & $1.0\times 10^8$    &  0.0117                  &                                  & $4.44\times10^4$  &             \\
\hline
\end{tabular}
\end{table}

\subsection{Packing fraction and coordination number}
\begin{figure*}
  \centering
  \includegraphics[width = 12.5 cm]{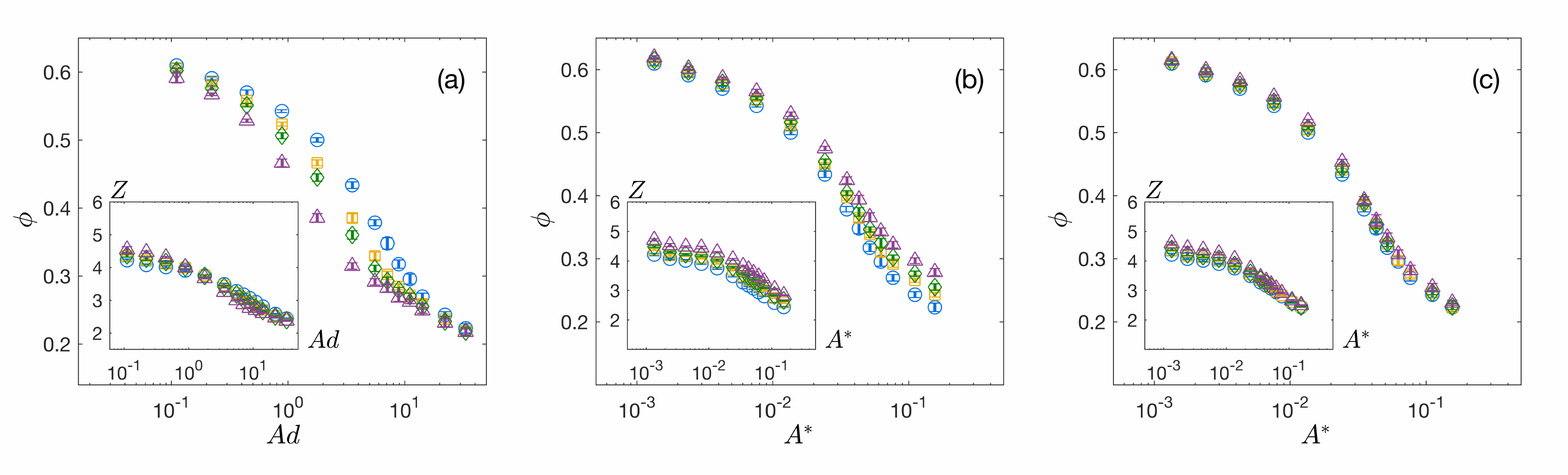}
  \caption{(a) Packing fraction $\phi$ and coordination number $Z$ as functions of adhesion parameter $Ad$ for packings with $El = 1.69 \times 10^6$ (case S, circles), $4.44\times 10^5$  (case A-1, squares), $2.22 \times 10^5$ (case A-2, diamonds) and $4.44\times10^4$ (case A-3, triangles). (b) $\phi$ and $Z$ as functions of parameter $A^* (= El^{-\frac{1}{3}}Ad^{\frac{5}{6}})$ for packings in series B, modified surface energy $\gamma_R = \chi^{2/5} \gamma_O$ are used and the rolling stiffness is calculated as $k_r = 12\pi \gamma_R R_{ij} \hat{a}_{ij}^{3/2}$. (c) $\phi$ and $Z$ as functions of parameter $A^*$ for packings in series C, modified surface energy are used and the rolling stiffness is calculated as $k_r = 12\pi \gamma_O R_{ij} \hat{a}_{ij}^{3/2}$}
  \label{fig_phiz}
\end{figure*}

Fig. \ref{fig_phiz} (a)-(c) show the variation of packing fraction $\phi$ and coordination number $Z$ as functions of adhesion parameter $Ad$ for series A, B and C, respectively. To avoid the wall effect, both $\phi$ and $Z$ are calculated from the middle part of the packing ($0.15h \leq x_p \leq 0.85h$, with $h$ as packing height). The blue circles in the three panels are data for cases S. From Fig. \ref{fig_phiz} (a), one can draw the conclusion that the reduction of the particle stiffness obviously decreases the packing fraction. This effect is more prominent in the range of moderate $Ad$. With a low adhesion number $Ad(<0.2)$ and a high $Ad (>10)$, the packing fraction converges to the random close packing limit (RCP) and {\color{black}{the adhesive loose packing limit (ALP)}}, respectively, \citep{LiuSM2015, LiuSM2017} and the difference in $\phi$ due to stiffness is totally prevented. In contrast to $\phi$, the coordination number $Z$ only has a weak dependence on particle stiffness. This interesting phenomenon may be understood through an analysis of the mechanical equilibrium of the packing. For a given contact network of a packing, mechanical stable condition is achieved with force- and torque-balance on all particles under the constraint $\mathcal{F} < \mathcal{F}_{crit}$ ($\mathcal{F}$ is $F_{ij}^S$, $M_{ij}^R$ or $M_{ij}^T$). According to Eq. (\ref{eqcrit_a}) and (\ref{eqcrit_c}), the critical value $\mathcal{F}_{crit}$ is independent of particle stiffness. A packing of harder particles can have each particle balanced at a similar coordination number as the packing of softer particles. However, before the final mechanical equilibrium is achieved, the kinetic energy of particles needs to be dissipated. A softer particle has a better capability of energy dissipation and is more likely to stick onto packed particles upon collisions, limiting its movement along the $x$ direction. In contrast, a particle with higher stiffness needs more times of collisions to be captured, which may lead to a compaction of the packing.

When the surface energy is modified according to Eq. (\ref{eqgamr}), both the packing fraction and the coordination number increases for particles with reduced stiffness and the case with higher reduced ratio $\chi$ tends to have a denser structure. Note that, in Fig. \ref{fig_phiz} (b) and \ref{fig_phiz} (c), we choose to use $A^* (= El^{-1/3}Ad^{5/6})$ instead of $Ad$ as the abscissa, because the modification of surface energy will shift the data points in $\phi - Ad$ plane. The difference in $\phi$ due to the reduction of the stiffness is, to some extent, balanced by the modification of the surface energy. However, there still remains considerable discrepancy. This discrepancy again can be attributed to the mechanical equilibrium: a reduction of surface energy causes the decrease of the critical value of rolling resistances $M_{ij,crit}^R$, which practically puts stricter constraints on the force- and torque-balance of particles. A packing with smaller $M_{ij,crit}^R$  generally needs more contacts to achieve mechanical equilibrium. These results, combined with the $e-v$ curves in Fig. \ref{fig_ev2}, indicate that an exactly same particle-particle normal collision behavior does not ensure the same results of packing structure.

In the last case, we modify the surface energy when calculating the normal forces but maintain the original value of the rolling stiffness $k_r$. As shown in Fig. \ref{fig_phiz} (c), the packings simulated with reduced stiffness well reproduce the structure, both $\phi$ and $Z$, of the original packings. This result confirms our statement that the critical value of rolling resistance strongly affects the mechanical equilibrium of a packing. In the framework of adhesive DEM with reduced stiffness, similarities in both particle-particle collision behavior and mechanical constraints are necessary to simulate a packing process. The friction coefficient $\mu$ is kept unchanged during the entire simulation since the value ($\mu = 0.3$) we use is large enough to ensure that rolling rather than sliding is the dominant mode of deformation of the packing. If the particles have a small friction coefficient, which is usually smaller than $0.05$, the sliding motion between contact particles will become non-negligible \citep{LiuPT2016, LiuSM2017}, and one should modify the critical value of $F_{crit}^S$ according to Eq. (\ref{eqmodslide})

\subsection{Local structure of packings}
\begin{figure}[h]
  \centering
  \includegraphics[width = 8.5 cm]{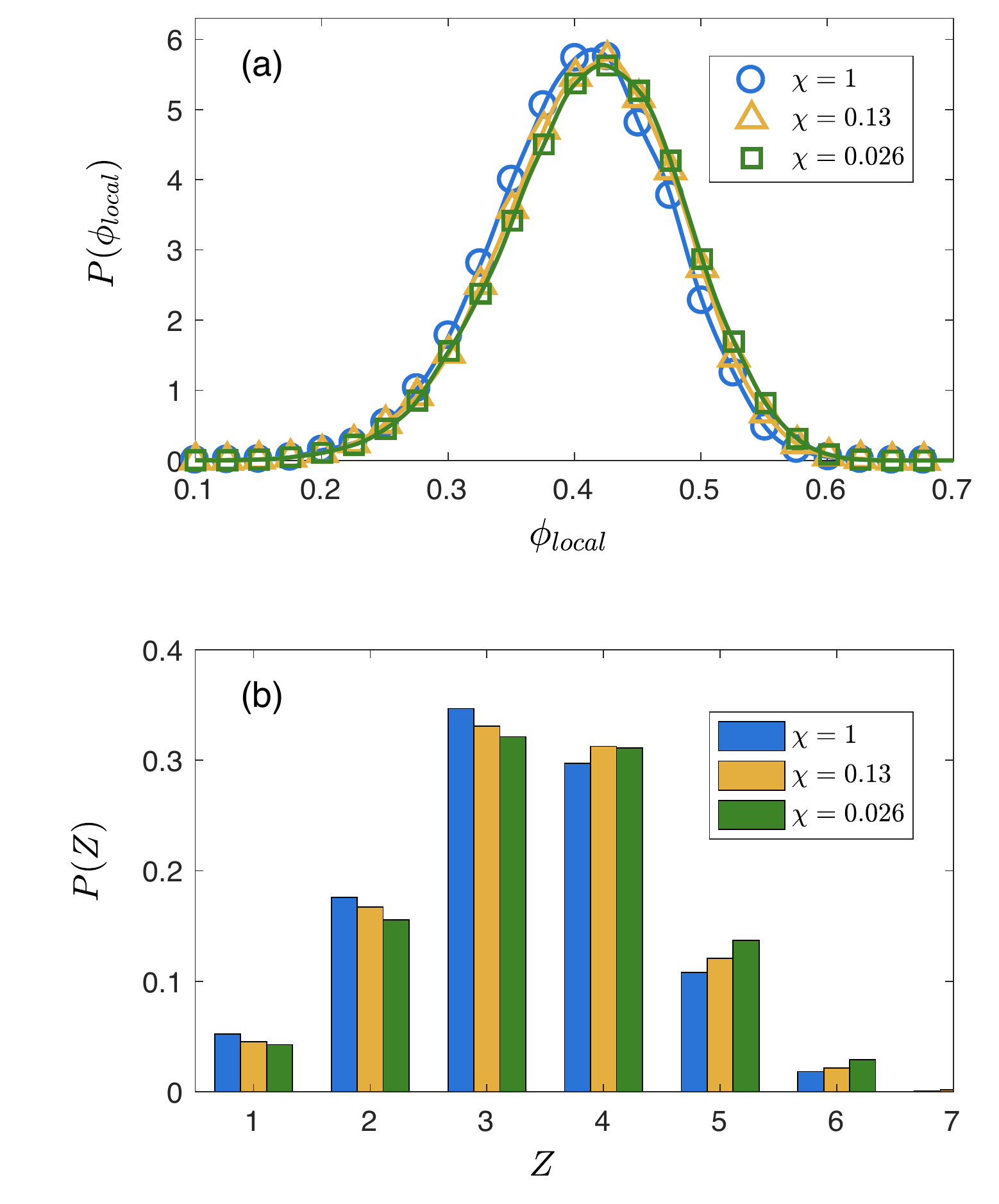}
  \caption{Distribution of local volume fraction $\phi_{local}$ and coordination number $Z$ of each particle. For each reduced ratio $\chi$, the PDF is averaged over 10 packings}
  \label{fig_dislocal}
\end{figure}

To further validate the fast adhesive DEM, we do statistics of the local structure of each particle inside a packing. We calculate the local packing fraction of each particle, which is expressed as
\begin{equation}
  \label{eq_vor}
  \phi_{local,i} = \frac{V_p}{V_{vor,i}},
\end{equation}
where $V_p$ is the volume of a particle and $V_{vor,i}$ is the volume of its Voronoi cell. Fig. \ref{fig_dislocal} shows the distributions of $\phi_{local,i}$ and coordination number $Z$ of each particle for the case S, C-2, and C-3 at $A^* = 0.035$. We choose this value because it locates in the transition region between RCP and ALP and packings in this region are more sensitive to the particle stiffness. There is a very good agreement between the PDFs obtained from original packings and from packings with reduced stiffness.

\subsection{Interparticle overlaps and normal forces}
One of the most important properties needs to be checked is the interparticle overlap, which usually puts a restriction on reducing the particle stiffness. Interparticle overlap significantly affects the heat or charge transfer between heated particles or charged particles \citep{Batchelor1977, MoyseyPT2005, JinJOE2017}. However, there is no universal criterion for choosing a limit of interparticle overlap. For example, it has been pointed out that the flow pattern on a bumpy inclined chute was not sensitive to stiffness when the interparticle overlap is smaller than $1\%$ of the particle diameter \citep{HanesPT2000}. In a measurement of the angle of repose, to retain the results, an overlap smaller than $0.34\%$ was suggested \citep{LommenPART2014}. Based on a broad review of different simulation tasks, \cite{PaulickPT2015} argued that, when the particle overlap is kept smaller than $1\%$ of the particle diameter, there would be no major change in the simulation result.

\begin{figure}
  \centering
  \includegraphics[width = 8.5 cm]{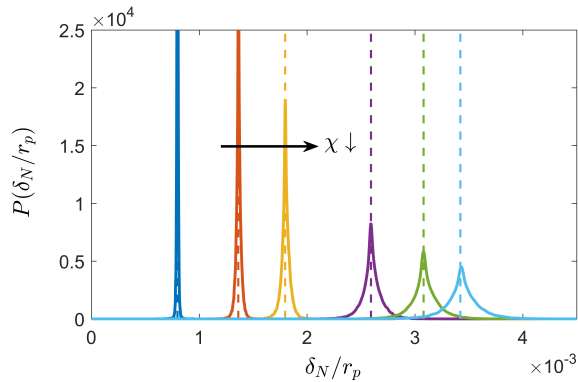}
  \caption{Distribution of scaled interparticle overlaps $\delta_N/r_p$. Curves from left to right, correspond to packings with $\chi = 1$, 0.26, 0.13, 0.053, 0.034, and 0.026, respectively. For each reduced ratio $\chi$, the PDF is averaged over 10 packings. The dashed lines indicate overlaps in the equilibrium state ($\delta_0/r_p$).}
  \label{fig_pdfdn}
\end{figure}

In Fig. \ref{fig_pdfdn}, we show the distributions of interparticle overlaps for Case S, C-1, C-2 and C-3 (corresponding to $\chi = 1$, 0.26, 0.132, and 0.026) at $A^* = 0.035$. Two extra reduced ratios, $\chi = 0.053$ and $0.034$, are also added. It is easy to understand that, as particle stiffness decreases, the distributions move to larger $\delta_N$. The interparticle overlaps are almost symmetrically distributed around the equilibrium value $\delta_0$ (indicated by the dashed vertical lines), which results from the balance between van der Waals attraction and the elastic repulsion. $\delta_0$ can be calculated written as:
\begin{equation}
  \label{eq_delta0}
  \delta_0 = 3.094\gamma^{\frac{2}{3}}R_{ij}^{\frac{1}{3}}E^{-\frac{2}{3}}.
\end{equation}
The symmetry in the distributions of interparticle overlap and normal force is a key feature of a static packing of strong adhesive particles \citep{LiuPT2016}. In Fig. \ref{fig_pdfdn}, the values of $\delta_O$ increases from $\delta_0/r_p = 0.08\%$ at $\chi = 1$ to $\delta_0/r_p = 0.34\%$ at $\chi = 0.026$, which is still within the range, $< 1\%$, suggested in \citep{PaulickPT2015}. From Eq. (\ref{eq_delta0}), one can easily evaluate the effect of reduction of stiffness on interparticle overlaps
\begin{equation}
  \label{eq_delta0red}
  \delta_{N,R} = \chi^{-2/5}\delta_{N,O}.
\end{equation}
This scaling allows users to determine a feasible amount of stiffness reduction once the constraint is put on the interparticle overlap.

\begin{figure}
  \centering
  \includegraphics[width = 8.5 cm]{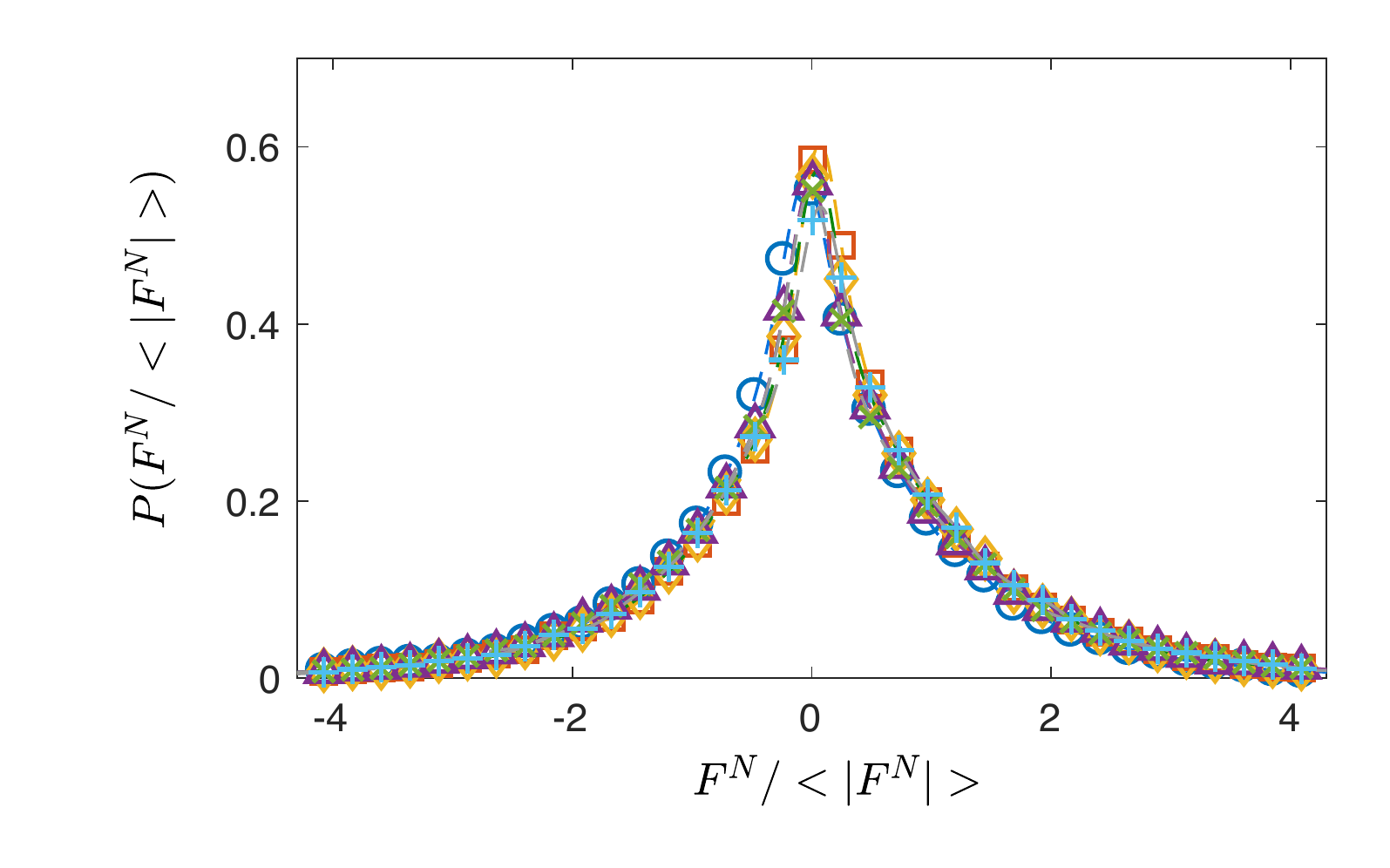}
  \caption{Distribution of scaled normal force $F^N / <|F^N|>$ for packings with $\chi = 1$ (circles), 0.26 (squares), 0.13 (diamonds), 0.053 (triangles), 0.034 (axes), and 0.026 (pluses). $<|F^N|>$ is the mean value of the magnitude of the normal force. Dashed lines are guides for the eye.}
  \label{fig_pdffn}
\end{figure}

It is also of great interest to know what the force distribution is like in packings, especially, in loose packings with adhesive particles. Here we measure the normal force of each contact in the same packings as those in Fig. \ref{fig_pdfdn}. As displayed in Fig. \ref{fig_pdffn}, the forces could be both attractive (negative $F^N$) and repulsive (positive $F^N$). After normalizing $F^N$ in each case with the corresponding mean value of its magnitude, $<|F^N|>$, distributions with different $\chi$ nicely collapse onto a single curve. The normalized distributions are almost symmetrical around $F^N/<|F^N|> = 0$, which is in good agreement with previous results on the packing of strong adhesive particles \cite{LiuPT2016}. The results again verify that the fast adhesive DEM with reduced particle stiffness can retain both the structural and mechanical properties of the contact network in a packing.

\begin{figure}
  \centering
  \includegraphics[width = 8.5 cm]{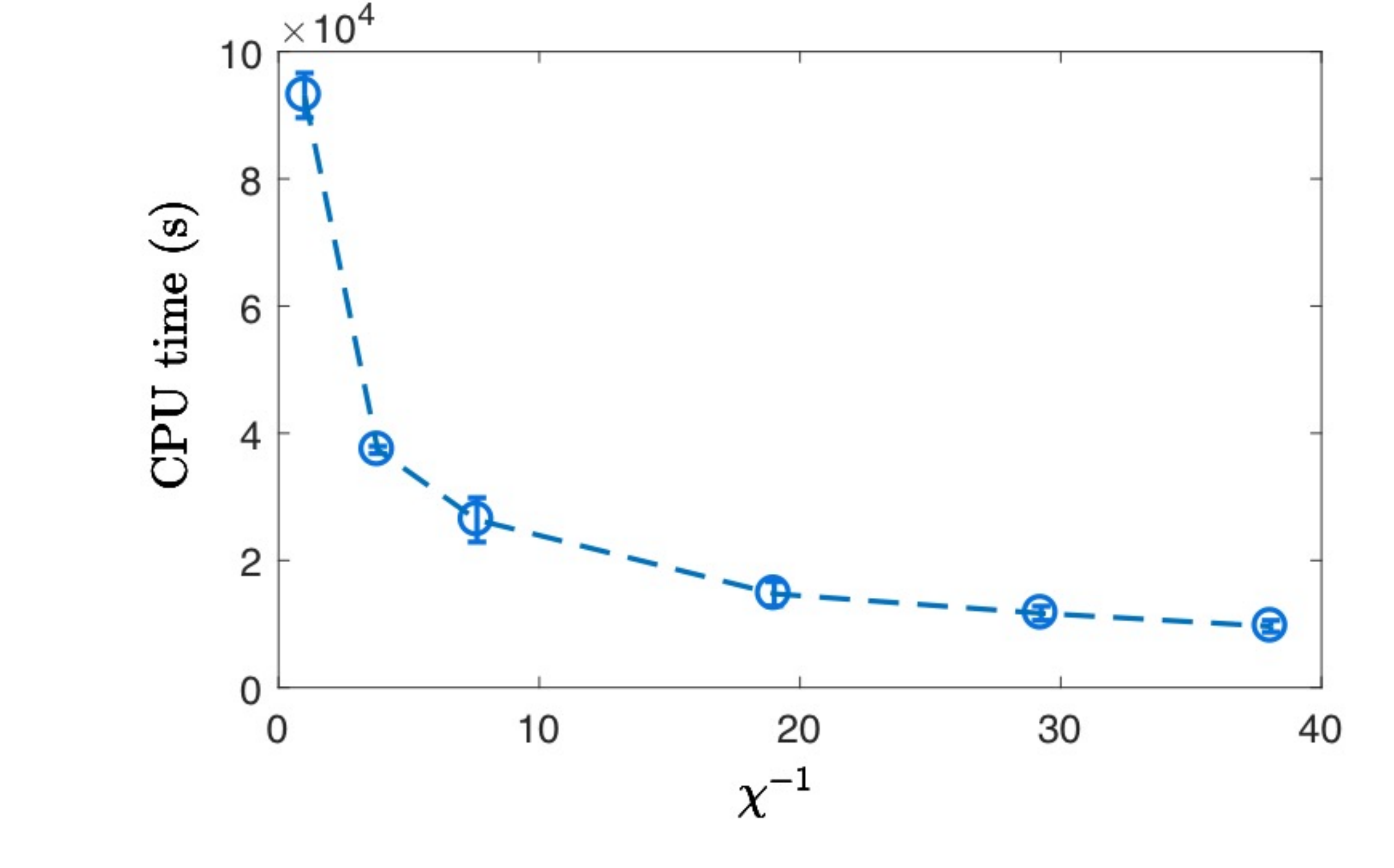}
  \caption{Timing (in seconds) results for $N = 2000$ packings with the reduced ratio $\chi (=E_R/E_O)$ of particle stiffness. Each data point is averaged over 10 runs.}
  \label{fig_time}
\end{figure}

At last, we report the timing results for the simulation of packings in Fig. \ref{fig_pdfdn}. Timing is measured on a computing node with 20-core Intel (R) Xeon (R) E5-2660 V3 running at 2.60 GHz and 128GB memory. The results in Fig. \ref{fig_time} indicate that reducing the particle stiffness by 1 or 2 orders of magnitude can shorten 5 times the computation time, however, further reduction in $\chi$ does not guarantee an obvious speedup. Combining the timing results and the scaling of interparticle overlap, we suggest that a reduction of stiffness by $1-2$ orders of magnitude can remarkably accelerate the simulation and retain both micro- and macroscopic properties of a static packing of adhesive particles.

\section{Discussion and Conclusions}
Based on the dimensionless equation describing the collision between a particle and a wall, we have been able to propose a scaling relationship to reduce particle's stiffness (i.e., particle's Young's modulus) and surface energy simultaneously. It allows one to use larger time steps to resolve the collision and ensure that the results stay the same. With a simple but indispensable modification of the rolling and sliding resistances, this accelerated JKR-based DEM can be feasibly applied to simulations of static packings of adhesive particles. Structural proprieties, including the overall packing fraction, the averaged coordination number and the distributions of local packing fraction and coordination number of each particle, are in good agreement with the packings simulated using the original parameters.

The current paper also presents a novel inversion method, which helps users to set the damping coefficient, particle stiffness and surface energy to reproduce a prescribed $e-v$ curve. This inversion method is different from previous calibration approaches, in which iterative procedure is normally used and the parameters are tuned to match the bulk response of the material to measured results \citep{Coetzee2017}. Compared with these calibration approaches, our approach uses practical formulas for a direct calculation avoiding complicated iteration process. Indeed, one can also determine the parameters based on a direct measurement of them at particle or contact level. However, experimental measurements are usually limited by particle sizes, and parameters like damping coefficient cannot be directly measured. Even if property values can be accurately measured, it is not guaranteed that the DEM model would show expected accuracy on the bulk level \citep{SimonsPE2015}. We suggest that the proposed inversion method should be used in combination with direct measuring approach. Parameters such as particle size and density are usually measured directly from experiments.

For the packing problem studied here, the final packing structures are essentially determined by particle-particle contact interactions, including (1) collisions between the incoming particles and the packed particles, which dissipate the kinetic energy of particles and (2) force- and torque-balance on all particles that ensures the mechanical stability of the packing. The reduced stiffness scaling and the inversion procedure proposed in this paper ensure the analogy in collisions (step (1)), and the modification of the resistances ensures the constraints on solving forces and torques for a given contact network (step (2)) are not affected by the reduction of stiffness. By now, we have neglected the effect of external forces, which actually exits in a variety of particulate systems. Typical external forces include fluid drag, gravity \citep{FanJFM2014}, electrostatic forces \citep{Ozel, ChenJFM2018}, etc. For particle systems with external forces, there should be an additional term, $\hat{F}_{ext}$, in Eq. (\ref{eqode2}):
\begin{equation}
  \label{eqpde3}
  \frac{\mathrm{d}^2\hat{\delta}}{\mathrm{d}\hat{t}^2} + A \hat{a}^{1/2}\frac{\mathrm{d}\hat{\delta}}{\mathrm{d}\hat{t}} + Bg(\hat{\delta}) - \hat{F}_{ext}= 0,
\end{equation}
where $\hat{F}_{ext} = F_{ext} \delta_C/mv_0^2$ is the non-dimensional form of the external force. As suggested in \citep{WashinoPT2018}, $F_{ext}$ should be modified to ensure $\hat{F}_{ext,R} = \hat{F}_{ext,O}$. However, we note that such modification may be reasonable only if the particle is in contact with other particles. For a free particle, one should use the original value of $F_{ext}$ to get a meaningful value of particle’s acceleration. A system containing both aggregates and individual free particles, such as a fluidized bed with gas flow rate around the minimum fluidization value, should be used to calibrate the fast adhesive DEM in the future.

\section*{Acknowledgements}
SQL acknowledges support from the National Fund for Distinguished Young Scholars of China (51725601) and National Key Research and Development Program of China (2016YFB0600602). We are grateful to Prof. Jeff Marshall at Vermont, Prof. Q. Yao and Prof. Y. Zhang at Tsinghua for their useful suggestions.

\section*{References}

\bibliography{ref_elastic_packing}

\end{document}